\input{epsf}
\documentstyle[aps,preprint]{revtex}
\begin{document}
\tightenlines

\title{Asymmetric dynamics and critical behavior\\
       in the Bak-Sneppen model}

\author{Guilherme J. M. Garcia$^*$ and Ronald Dickman$^\dagger$\\
 {\small Departamento de F\'\i sica, Instituto de Ci\^encias Exatas}\\
 {\small Universidade Federal de Minas Gerais, Caixa Postal 702}\\
 {\small CEP 30123-970, Belo Horizonte - Minas Gerais, Brazil} }

\date{\today}

\maketitle
\vskip 0.5truecm

\begin{abstract}

We investigate, using mean-field theory and simulation, the effect of asymmetry on
the critical behavior and probability density of Bak-Sneppen models. Two kinds of
anisotropy are investigated: (i) different numbers of sites to the left and right of the 
central (minimum) site are updated and (ii) sites to the left and right of the 
central site are renewed in different ways. Of particular interest is the crossover 
from symmetric to asymmetric scaling for weakly asymmetric dynamics, and the collapse 
of data with different numbers of updated sites but the same degree of asymmetry. 
All non-symmetric rules studied fall, independent of the degree of asymmetry, 
in the same universality class. Conversely, symmetric variants reproduce the exponents
of the original model.  Our results confirm the existence of two symmetry-based 
universality classes for extremal dynamics.

\vspace{1em}
\vspace{2em}

\noindent$^*$ Electronic address: gjmg@fisica.ufmg.br\\
$^\dagger$ Electronic address: dickman@fisica.ufmg.br

\end{abstract}

\newpage

\section{INTRODUCTION}

Nature exhibits scale invariance in a variety of settings. 
Such behavior is often associated with
power law distributions of the phenomenon of interest, 
for example earthquake sizes \cite{terremotos,BTW},
rain intensities and drought durations \cite{lavergnat,Peters,rain}
and physiological and morphological quantities \cite{Scaling,West}.
In recent decades, physicists have sought the physical origin 
of such laws \cite{BTW,rain,West}.
A concept introduced to partially explain the ubiquity of scale-invariant
phenomena in nature is so-called {\sl self-organized criticality} or SOC
\cite{BTW,BS}.

The Bak-Sneppen (BS) model \cite{BS,flyvbjerg} was proposed to explain mass extinctions
observed in the fossil record, and has attracted much attention as a prototype
of SOC under extremal dynamics \cite{Dickman et al:2000}.
The model has been studied through various approaches,
including simulation \cite{Grassberger:1995,Paczuski,Rios:1998,Boettcher:2000},
theoretical analysis \cite{Meester:2002,Li and Cai:2000,Dorogovtsev:2000},
probabilistic analysis (run time statistics) \cite{Caldarelli:2002,Felici:2001},
renormalization group \cite{Marsili:1994,Mikeska:1997},
field theory \cite{Paczuski:1994}
and mean-field theory \cite{flyvbjerg,Dickman et al:2000,deBoer:1994,deBoer:1995,Pismak:1997,head,Garcia and Dickman:2003};
it was recently adapted to model experimental data on bacterial populations 
\cite{Donangelo and Fort:2002,Bose:2001}.  Applications of the model in evolution
studies are reviewed in \cite{drossel02}.
Asymmetric versions of the model were studied in Refs. \cite{Head and Rodgers:1998} 
and \cite{Maslov:1998}.
In this paper, we study how varying the degree of symmetry affects the
stationary distribution and the critical behavior, using mean-field theory
and numerical simulation.

In the evolutionary interpretation of the BS model, each site $i$
represents a biological species, and bears a real-valued variable
$x_i$ representing its ``fitness".
The larger $x_i$, the better adapted this species is to its
environment and so the more likely it is to survive. At each time step,
the site with the smallest $x_i$ and its nearest neighbors are
replaced with randomly chosen values. The replacement of $x_i$ represents extinction
of the less-fit species and the appearance of a new one, while the substitution of the
neighboring variables with new random values may be interpreted
as a sudden unpredictable change in fitness when an interdependent 
species goes extinct and a new species colonizes its niche.
Selection, at each step, of the global minimum of the $\{x_i\}$
(``extremal dynamics") represents a
highly nonlocal process, and would appear to require an external 
agent with complete information regarding the state of the system at each
moment.

Due to the extremal dynamics, this system exhibits scale-invariance in
the stationary state, in which several quantities display power-law behavior
\cite{BS}. Simulations show that the stationary distribution of barriers
follows a step function, being zero (in the infinite-size limit) for
$x < x^* \simeq 0.66702(3)$ \cite{Paczuski}.  Relaxing the extremal
condition leads to a smooth probability density and loss of scale invariance 
\cite{Dickman et al:2000,head,Garcia and Dickman:2003}.
Datta et. al have also shown that the scaling behavior is sensitive
to the number of sites updated at each step (i.e., updating
only the minimum site,
or the minimum and next-to-minimum as well) \cite{datta}.

In this paper we investigate the effect of symmetry in the updating rule.
Section II introduces the models, which are then analyzed using mean-field
like approaches in Sec. III.  In Sec. IV we present simulation results;
our conclusions are summarized in Sec. V.

\section{MODELS}

The Bak-Sneppen model \cite{BS} is defined as follows.
Consider a $d$-dimensional
lattice with $L^d$ sites and periodic boundaries. In the evolutionary
interpretation of the model, each site $i$ represents a species, and
bears a real-valued variable $x_i(t)$ representing its ``fitness'' to survive,
so that $x_i(t)$ may be termed a ``barrier to extinction".
The initial values of the barriers are independently drawn from a uniform
distribution on the interval [0,1). 
At time 1, the site $m$ bearing the minimum of all the numbers
$\{x_i(0) \}$ is identified, and it, along with its $2d$ nearest neighbors,
are given new random values, again drawn independently from
the interval [0,1). (In the one dimensional case considered here
this amounts to: $x_m(1) = \eta$, $x_{m + 1}(1) = \eta'$, and $x_{m - 1}(1) = \eta''$,
where $\eta$, $\eta'$, and $\eta''$ are independent and
uniformly distributed on [0,1); for $|j-m|>1$, $x_j(1) = x_j(0)$.) 
At step 2 this process is repeated, with $m$ representing the site with the
global minimum of the variables $\{x_i(1)\}$, and so on.

We now define several one-dimensional variants of the BS model, which differ 
from the original in the number and/or position of neighbors which are updated 
at each time step,
or in the way that the barriers $x_i$ evolve. In the `generalized' or `BSab' variant,
we replace the site $m$ bearing the minimum of the $\{x_i\}$ plus $a$ neighbors on the
left side and $b$ neighbors on the right with independent random numbers. 
If $a=b=1$ we recover the original model (BS11); if $a=0$ and $b=1$ we have
the anisotropic BS model (BS01) studied in 
\cite{Head and Rodgers:1998,Maslov:1998}; if $a\neq b$ we obtain modified BS
models with asymmetric dynamics. 
These BSab variants of the Bak-Sneppen model were also studied in \cite{Head and Rodgers:1998}.
In the second variant, the site $M$ bearing the {\sl maximum} of the $\{x_i\}$
and its two nearest neighbors are updated according to the rules:
$x_{M}(t+1) = \eta$, $x_{M'}(t+1) =\eta '$ and $x_{M''}(t+1) = [x_{M''}(t)]^2$.
Here $M'=M+\sigma$ and $M''=M-\sigma$, where $\sigma$ is $+1$ with probability $p$ 
and $-1$ otherwise.
We shall refer to this variant as the peripheral square model with variable anisotropy.
(For $p=1/2$ this is the peripheral square model studied in \cite{Garcia and Dickman:2003}.)

The motivation for studying these variants is threefold. First it is of interest to
examine the effect of 
(i) the symmetry of the dynamics and
(ii) replacement of barriers with a deterministic function $f(x)$ instead of random 
numbers, on the critical behavior of the model. Secondly, we study corrections to
the power-laws, such as finite-size effects, and find that these corrections are
large for small asymmetries in agreement with \cite{Head and Rodgers:1998}.
Finally, since the precise form of the dynamics in a specific setting
(e.g., evolution) is generally unknown, and probably is quite different from
that of the original model, it is of interest to test the robustness
of the results reported for the original model.  It is even possible that
some of the variants considered approximate a given evolutionary process
more closely than the original.   In particular, if certain pairs of
species ($i$ and $j$, say) stand in a predator-prey relationship, one 
would not expect the extinction of $i$ to have the same effect on $j$
as the extinction of $j$ has on $i$; in this situation an
asymmetric interaction appears more reasonable.

\section{MEAN-FIELD THEORY}

We develop a mean-field theory for the BSab variants, along the lines 
of Refs. \cite{Dickman et al:2000,head}. To begin, we introduce a flipping 
rate of $\Gamma e–^{-\beta x_i}$ at site $i$, where $\Gamma^{-1}$ is a 
characteristic time, irrelevant to stationary properties, and which we 
set equal to one ($\Gamma$ = 1).
This regularized system is the `finite-temperature' model 
\cite{Dickman et al:2000,head,Garcia and Dickman:2003}, in which all sites 
have a nonzero 
probability of being updated at any time, in contrast to the extremal dynamics
of the original model. The extremal condition is recovered in the zero-temperature
limit, $\beta \to \infty$; a regularized flipping rate
facilitates analysis of the model.

The probability density $p(x,t)$ satisfies:

\begin{eqnarray}
\nonumber
{dp(x,t)\over{dt}} &=& -e^{-\beta x} p(x,t) 
-\sum_{j=1}^a \int_0^1 e^{-\beta y} p_j(x,y,t)dy
-\sum_{j=1}^b \int_0^1 e^{-\beta y} p_j(x,y,t)dy
\\
&+& n\int_0^1 e^{-\beta y} p(y,t)dy ,
\label{eq:1}
\end{eqnarray}

\noindent where $n=a+b+1$ is the number of sites that are updated 
at each step,
$p_j(x,y,t)$ is the joint density for sites $0$ and $j$ and
$p(y,t)$ is the one-site marginal density. 
(We assume translation and reflection invariance, which is expected to hold
at any time, if the initial distribution possesses these properties.)
Invoking the mean-field factorization
$p_j(x,y,t)=p(x,t)p(y,t)$, we find:

\begin{equation}
{dp(x,t)\over{dt}} = -p(x,t)[e^{-\beta x} + (n-1)I(\beta)] +nI(\beta) ~~,
\label{eq:2}
\end{equation}

\noindent where

\begin{equation}
I(\beta) \equiv \int_0^1 e^{-\beta y} p(y,t)dy 
\label{eq:3}
\end{equation}
represents the overall flipping rate.
In the stationary state we have

\begin{equation}
p_{st}(x) = \frac{nI}{(n-1)I+ e^{-\beta x}} ~~ .
\label{pss}
\end{equation}

\noindent Multiplying by $e^{-\beta x}$ and integrating over the 
range of $x$, we find $I(\beta) = (e^{(n-1)\beta/n}-1)/[(n-1)e^\beta(1-e^{-\beta/n})]$
and thus
\begin{equation}
p_{st}(x) = {n\over{(n-1)}}
{{1-e^{-(n-1)\beta/n}}\over{1-e^{-(n-1)\beta/n}+e^{-\beta x}(e^{\beta/n}-1)}}
~~.
\label{eq:4}
\end{equation}
In the limit $\beta \rightarrow \infty$ this solution becomes a  step function:

\begin{equation}
p_{st}(x) = {n\over{(n-1)}} \Theta(x-1/n)\Theta(1-x) ~~ .
\label{eq:5}
\end{equation}

\noindent Thus, the mean-field approach predicts a step-function singularity for
the probability density with the critical barrier at $x^* = 1/n$. This result is
independent of the symmetry of the dynamics, and we conclude that the mean-field
approximation at the level of the one-site marginal density is 
insensitive to differences in symmetry.

Define the anisotropy coefficient
\begin{equation}
k_a={|a-b|\over n ~~}.
\label{defka}
\end{equation}

\noindent (Note that $k_a=0$ for symmetric dynamics.)
The mean-field threshold $x^* = 1/n$ can be written as

\begin{equation}
x^* = {1\over{(2a+1)}}(1-k_a) ~~ {\rm for} ~ b>a.
\end{equation}

\noindent For fixed $a$, mean-field theory predicts that the threshold $x^*$
varies linearly with the anisotropy coefficient $k_a$.

\section{SIMULATIONS}

\subsection{Threshold values}

We study the generalized model for various values of $a$ and $b$,
corresponding to $k_a$ in the range 0$-$0.857.
We estimate the probability density $p(x)$ on the basis of a histogram of
barrier frequencies, dividing [0,1] into 100 subintervals.
After the system (a ring of $N=2000$ sites) relaxes to the stationary 
state\footnote{Typically, the system can be considered to be in the stationary state
after $\approx 10^7$ time steps.},
histograms are accumulated at intervals of $N$ time steps until a total of
$10^8$ steps are performed.

The threshold $x^*_N$ for a finite system can be calculated
as follows \cite{Garcia and Dickman:2}. In the limit $N\to \infty$, the 
probability density $p(x)$ is a step function, with $p(x)=0$ for $x<x^*$
and $p(x)=C$ for $x>x^*$, where $C$ is a constant. Normalization implies
that $C(1-x^*)=1$. This suggests that $x^*_N=1-1/C_N$ can be regarded as
the threshold of a finite system. In ref. \cite{Garcia and Dickman:2}
we performed simulations for various systems sizes and found that
$x^*_N-x^*_\infty = k N^{-1/\nu}$, with $x^*_\infty=0.6672(2)$ and 
$\nu=1.40(1)$ in the original model (or BS11) and $x^*_\infty=0.7240(1)$
and $\nu=1.58(1)$ in the anisotropic BS model (or BS01). 
Typically $x^*_N-x^*_\infty \approx 0.006$ for $N=2000$ and, therefore, 
the threshold $x^*_{N=2000}$ is sufficiently accurate for our
present purpose, which is to investigate the effect of anisotropy on the
threshold values.

Simulation confirm that the stationary probability density is
a step function, in agreement with MFT.
Fig. 1 shows $p(x)$ for all variants having $n=5$. 
The threshold values, listed in Table I, are however always larger than
that predicted by MFT, $x^*=1/n$.  MFT is 
exact for the random-neighbor version, in which all
sites are considered neighbors, and in which ($n-1$) randomly
selected sites are updated in addition to the site with the minimum
$x_i$.  There is better agreement between the 
MFT prediction for $x^*$ and the simulation result
for larger anisotropies (with $n$ constant),
and for larger $n$ (with fixed $k_a$).  
MFT predicts that $x^* = x^*(n)$, whereas in fact, for fixed $n$, 
the threshold is {\it smaller}, the larger the anisotropy. 
This may be understood in general terms as follows.  With $n$ fixed,
the larger $|a-b|$, the greater is the distance between updated 
sites and the minimum site.  This makes the dynamics resemble
the mean-field (random neighbor) case more closely, causing the
threshold to tend toward the MFT value $1/n$.  Increasing $n$
(with $k_a$ fixed) should have a similar effect.

Figure 2 shows that if we fix $a$ and vary $b$, the threshold decreases
linearly with $k_a$, as predicted by MFT, except for $a=0$.
The actual threshold values are, however, quite different from MFT
results. For instance for $a=1$, MFT predicts $x^*=(1/3)(1-k_a)$,
while a fit to the simulation data yields $x^*=0.664(3)-0.668(7)k_a$.

\subsection{Critical exponents}

Several quantities display power-law behavior in the Bak-Sneppen model,
and can be used to characterize the associated universality class
\cite{BS,Grassberger:1995,Head and Rodgers:1998}.
In particular, we study the probability $P_J(r)$ that
successive updated sites are separated by a distance $r$.
In the original model, one finds \cite{Head and Rodgers:1998}  

\begin{equation}
P_J(r) \sim r^{-\pi}, ~~ {\rm with} ~~ \pi={\pi}_S=3.23(2)
\end{equation}

\noindent (figures in parentheses denote uncertainties).
On the other hand, in the anisotropic BS model (BS01),
$\pi={\pi}_A=2.401(2)$ \cite{Maslov:1998}.

Based on simulations of the generalized models using
$7\times 10^8$ time steps on lattices of 32000 sites we
find that all variants with asymmetric dynamics belong 
to the universality class of the anisotropic BS model.
Fig. 3 displays $P_J(r)$ for variants
with different anisotropies. The corrections to the power
laws are large for small asymmetries and therefore
the asymptotic behavior is not observed in small lattices
in these cases, because large values of $r$ are not reached.

We find that $P_J(r)$ can be fitted with an expression representing
a crossover between symmetric scaling at small $r$ and asymmetric
scaling at large $r$,

\begin{equation}
P_J(r) = A [r^{-\pi_A} + (N-r)^{-\pi_A} ] + B [r^{-\pi_S}  + (N-r)^{-\pi_S}]~~,
\label{expr}
\end{equation}

\noindent where $A$ and $B$ are constants and $\pi_A$ and $\pi_S$ are,
respectively, the exponents of the anisotropic and the original models
quoted above. The `mirror' structure of Eq. (\ref{expr}) arises due to
the periodic boundary conditions, which imply that $P_J(r)= P_J(N-r)$.
Eq. (\ref{expr}) provides a good fit to the simulation data for $r>10^2$
for all generalized models. Since $\pi_S \approx \pi_A+1$, the expression
$P_J(r) = A r^{-\pi_A} (1+B/r) + A (N-r)^{-\pi_A}(1+B/(N-r))$ also
fits the data well. Nevertheless, studies of weakly asymmetric models
(e.g., BS(20)(21), for which $k_a=0.024$) indicate that the slope 
of $P_J(r)$ (on log scales)
approaches $\pi_S$, as opposed to $\pi_A + 1$, 
before the asymptotic behavior is attained.

The functions $P_J(r)$ for variants with the same anisotropy
can be collapsed onto a master curve

\begin{equation}
P_{collapse}(k_a,r')=n\ P_J(r/n) ~~,
\label{colapso}
\end{equation}

\noindent where $r'=r/n$.
Figure \ref{fig4} illustrates this collapse for five variants with $k_a=1/6$.
A simple argument provides an intuitive understanding of this scaling function.
Suppose site $m$ is the extremal site at time $t$. Since
renewed sites receive independent random numbers, all sites in the set 
${\cal N}(t) \equiv \{m(t)-a, m(t)-a+1, ..., m(t)+b\}$
have the same probability of being the extremum at time $t+1$,
which implies (with $a \leq b$)
\begin{equation}
2 P_J(0) = P_J(1) = P_J(2)=...= P_J(a),
\label{plat1}
\end{equation}
so that $P_J(r)$ exhibits a plateau for $r = 1,...,a$.

The probability of event $m(t+1) \in {\cal N}(t)$, i.e.,
the extremal site belongs to the group of sites updated at 
the last time step, can be
estimated as follows.  We assume that, with a probability 
 $\approx 1$, all sites outside of ${\cal N}(t)$ have $x_i > x^*$.
Then the probability that $m(t+1) \in {\cal N}(t)$
is  $1 - (1-x^*)^n$.  In MFT we then have
$P(1)={2\over n} [1-(1-x^*)^n]\propto {1\over n}$, consistent
with the prefactor $n$ in Eq. (\ref{colapso}).
(Observe that for the original BS model, $1 - (1-x^*)^n \simeq 0.96$.)

If $b > a$, then 
\begin{equation}
P_J(a+1) \simeq P_J(a+2) \simeq \cdots \simeq P_J(b) 
\simeq \frac{1}{2} P_J(1)
\label{plat2}
\end{equation}
where the symbol `$\simeq$' is used because there
is a small probability of one of the distances $r = a+1,...,b$
being the extremum distance even if $m(t+1) \notin {\cal N}(t)$.
This explains the second plateau seen in Fig. 4.  For fixed $k_a$,
$a = [n(1-k_a) - 1]/2$ and $b = [n(1+k_a) -1]/2$ are essentially
proportional to $n$, so that the rescaling of the argument in
Eq. (\ref{colapso}) collapses the plateaus.  For $r > b$
the probability $P_J$ rapidly approaches a power law, which is also
invariant under the rescaling of Eq. (\ref{colapso}).
Although the foregoing arguments are approximate, they provide 
an intuitive basis for the collapse seen in Fig. 4.

Another quantity that obeys a power law in the Bak-Sneppen model is $P_r(\tau )$,
the probability that, in the stationary state,  the global minimum site at time 
$t$ was the extremal site most recently at time $t-\tau$. ($\tau $ is the
`return time.') We simulate the BSab variants on a lattice of $16000$ sites, 
using $7\times 10^8$ time steps. The results are quite similar to those 
for the probability $P_J(r)$.  For weak anisotropies, the asymptotic 
behavior is observed only for large $\tau$, as shown in Fig. \ref{fig5}. 
In the limit $\tau \to \infty$, $P_r(\tau )\propto \tau^{-b}$, where $b=1.40(1)$ for
all anisotropic variants and $b=1.58(1)$ for all isotropic cases.

\subsection{Peripheral square model with variable anisotropy}

The peripheral square model with variable anisotropy, defined in Sec. II, is a
generalization of the model studied in \cite{Garcia and Dickman:2003},
where the anisotropy now depends on a parameter $p$. We simulated this model
(with $N=2000, 4000$ and $8000$ sites) for $p$ varying from the isotropic
value ($p=0.5$) to the maximally anisotropic one ($p=1.0$). Our analysis shows
that, in the limit $L\to \infty$, the stationary probability density consists
of two regions (see Fig. \ref{fig6}): $p_{st}(x)=0$ for $x>x^*$ and 
$p_{st}(x)\ne 0$ for $x<x^*$. (The rounding in the threshold region is
simply a finite-size effect.) This behavior is in agreement with
Ref. \cite{Garcia and Dickman:2003}, where we found that 
renewing the barriers via $x' = x^\alpha$, with $\alpha=1/2$ and 2,
leads to a diversity of distributions with a discontinuity at
the threshold. 
The divergence of $p_{st}$ as $x \to 0$ in the present case can be understood
on the basis of the mean-field theory developed in \cite{Garcia and Dickman:2003}.
We conclude that the step-function distribution
of the original BS model is not universal for self-organized criticality
under extremal dynamics. In contrast, our results suggest that the 
step-singularity is universal.

Figure \ref{fig6} suggests that, in the limit $x\to 0$, $p_{st}(x)$ does not
depend on the parameter $p$. 
Over restricted intervals $p_{st}$ appears to follow a power law.  For example,
for $x\in [10^{-6},10^{-4}]$, $p_{st}$ is well
described by a power law with an exponent that increases from 0.82 for $p=0.5$ to
0.85 for $p=1.0$.  On the other hand, the mean-field solution, which appears to
capture the qualitative behavior, does not follow a power law, diverging
instead as $\sum_{n=0}^{\infty} 4^{-n} x^{2^{-n}-1}$ 
in the limit $x\to 0$.  (We further note that although the MF solution accompanies
the general trend of the data, it exhibits a series of step singularities in addition to the
jump at $x^*$.  Several step discontinuities are in fact observed in 
$p_{st}$ in simulations of the
random-neighbor version \cite{Garcia and Dickman:2003}, although not in the nearest-neighbor
version.)

The dependence of the threshold $x^*$ on $p$ is shown in Fig. \ref{fig7}.
Note that augmenting the anisotropy, $x^*$ increases almost linearly.
Therefore, similarly to the BSab variant, the effect of increasing the anisotropy,
while maintaining the number of updated sites constant, is to decrease 
the interval on which $p(x)=0$.

Figure \ref{fig8} shows the probability density $P_J(r)$ for various values of $p$.
While the isotropic case ($p=0.5$) preserves the exponent of the original BS model,
all cases for $p\neq 0.5$ exhibit the exponent of the anisotropic BS model. 
In the BSab variant studied above, asymmetry was due to different numbers of
sites being renewed in each side of the extremal site. In the present case we have a
different kind of asymmetry, namely, sites are updated in different ways on each
side of the extremal one. We nevertheless find the
same critical exponents as for the anisotropic BS model. This strengthens the
evidence for the existence of two symmetry-based universality classes for 
models under extremal dynamics.

\section{CONCLUSIONS}

We perform a detailed investigation of the effect of symmetry on the 
scaling behavior of the Bak-Sneppen model.
All dynamics which preserve the reflection symmetry of the original 
BS model possess the same critical exponents as the
original model, while asymmetric dynamics lead to the exponents of the anisotropic BS model.
Therefore, our work reinforces the evidence for two symmetry-based universality
classes \cite{Head and Rodgers:1998}.

In order to obtain these results, we define modified BS models, and study 
them via mean-field theory and simulation. In the generalized BS model 
(or BSab variant), the degree of asymmetry is quantified
by the anisotropy coefficient $k_a$. Mean-field theory provides poor 
predictions for the threshold $x^*$, but correctly predicts 
(i) that the stationary probability density is a step function, and
(ii) that $x^*$ varies linearly with the number $n$ of updated sites, 
as found in simulations on a one-dimensional lattice.
The simulations also lead to several conclusions that go beyond MFT analysis:
(i) the threshold value decreases as the degree of anisotropy is increased;
(ii) for weak anisotropy we observe a crossover between symmetric
and asymmetric scaling; (iii) for fixed anisotropy we find a
scaling collapse of the probability $P_J(r)$ that successive minimum sites
have a separation $r$.

In the peripheral square model with variable anisotropy, the degree of asymmetry
is quantified by the parameter $p$ that controls on which side of the extremal 
site the neighbor is updated differently. Although this model
is partially deterministic and has a different kind of asymmetry,
we again encounter two symmetry-based universality classes. Moreover, we find that
increasing the asymmetry, the region where $p(x)=0$ is reduced, as in the
the BSab variant.

These results, together with evidence for
finite-size scaling \cite{Garcia and Dickman:2}, and connections
with directed percolation \cite{Paczuski:1994}, indicate that scaling
in the BS model partakes of many of the characteristics associated
with critical phenomena, both in and out of equilibrium.  The
particular distinguishing features of the Bak-Sneppen model and
its variants appear to be associated with extremal dynamics.

\bigskip \bigskip
\noindent {\bf ACKNOWLEDGMENTS}
\medskip

\noindent We thank CNPq, CAPES and FAPEMIG, Brazil, for financial support.

\begin{table}
\caption{Threshold ($x^*_N$) for systems of $N=2000$ sites for 
generalized BSab models
with various anisotropy coefficients ($k_a$). 
Figures in parentheses denote uncertainties.}
\begin{center}
\begin{tabular}{|c|c|c|}
MODEL  &   $x^*_{N=2000}$    &   $k_a$   \\
\hline
 BS01  &  0.717(2) &  1/2 \\
 BS02  &  0.531(2) &  2/3 \\
 BS03  &  0.417(2) &  3/4 \\
 BS04  &  0.342(2) &  4/5 \\
 BS05  &  0.290(2) &  5/6 \\
 BS06  &  0.251(2) &  6/7 \\
 BS11  &  0.661(9) &  0   \\
 BS12  &  0.501(3) &  1/4 \\
 BS13  &  0.398(2) &  2/5 \\
 BS14  &  0.329(3) &  1/2 \\
 BS15  &  0.280(2) &  4/7 \\
 BS22  &  0.466(9) &  0  \\
 BS23  &  0.381(3) & 1/6 \\ 
 BS24  &  0.317(3) & 2/7 \\
 BS25  &  0.271(2) & 3/8 \\
 BS26  &  0.237(2) & 4/9 \\
 BS27  &  0.210(2) & 1/2 \\
 BS33  &  0.36(1)  &  0  \\
\end{tabular}
\end{center}
\end{table}

\begin{figure}
\epsfysize=6cm
\epsfxsize=7.2cm
\centerline{
\epsfbox{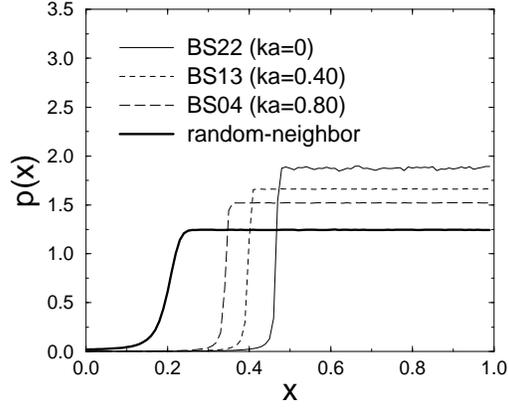}}
\caption{Stationary probability densities for the BSab with $n=5$.}
\label{fig1}
\end{figure}

\begin{figure}
\epsfysize=6cm
\epsfxsize=7.2cm
\centerline{
\epsfbox{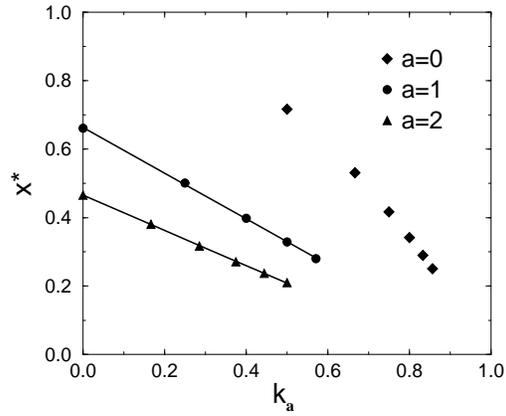}}
\caption{Threshold values as a function of the anisotropy coefficient $k_a$.
We fix $a$ and change $b$ to vary $k_a$. Further explanation in text.}
\label{fig2}
\end{figure}

\begin{figure}
\epsfysize=6cm
\epsfxsize=7.2cm
\centerline{
\epsfbox{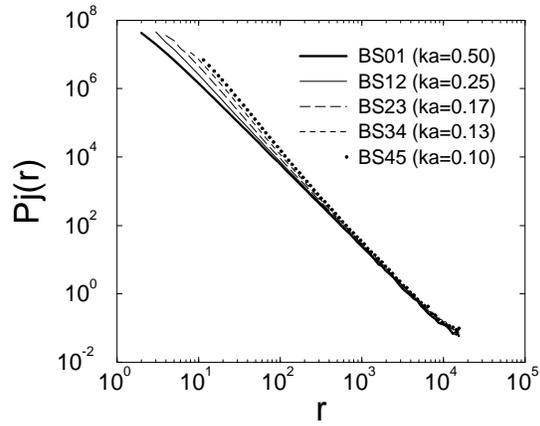}}
\caption{Stationary probability $P_J(r)$ for various BSab variants.}
\label{fig3}
\end{figure}

\begin{figure}
\epsfysize=6cm
\epsfxsize=7.2cm
\centerline{
\epsfbox{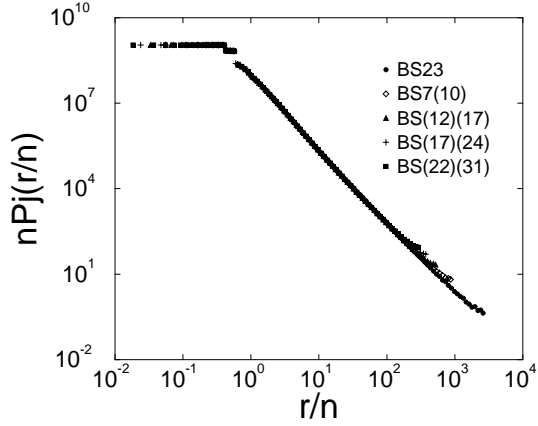}}
\caption{Scaling collapse of $P_J(r)$ for variants with $k_a=1/6$, 
see Eq. (\ref{colapso}).}
\label{fig4}
\end{figure}

\begin{figure}
\epsfysize=6cm
\epsfxsize=7.2cm
\centerline{
\epsfbox{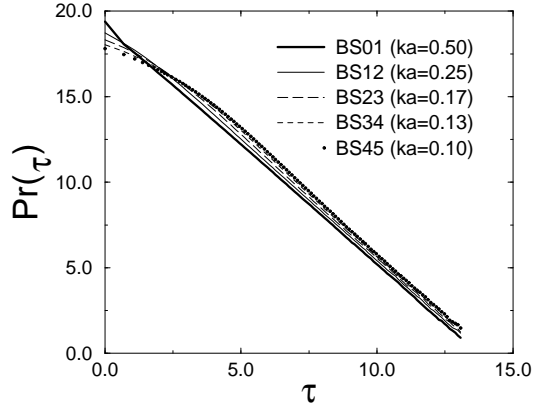}}
\caption{Stationary probability $P_r(\tau )$ for BSab variants.}
\label{fig5}
\end{figure}

\begin{figure}
\epsfysize=6cm
\epsfxsize=7.2cm
\centerline{
\epsfbox{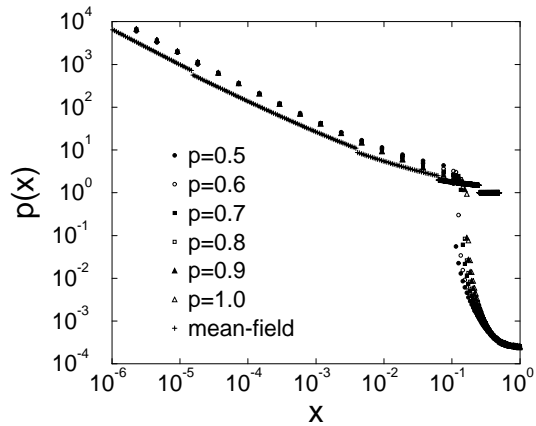}}
\caption{Stationary probability density $p(x)$ given by mean-field theory and simulation
($N=8000$ sites). The parameter $p$ varies from the symmetric value (p=0.5) to the maximaly 
asymmetric one (p=1.0).}
\label{fig6}
\end{figure}

\begin{figure}
\epsfysize=6cm
\epsfxsize=7.2cm
\centerline{
\epsfbox{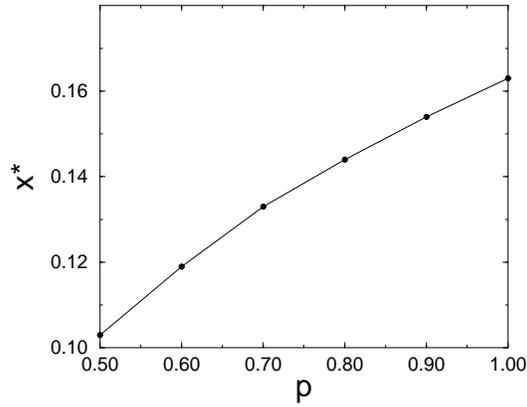}}
\caption{Thresholds of the peripheral square models as a function of the parameter $p$.}
\label{fig7}
\end{figure}

\begin{figure}
\epsfysize=6cm
\epsfxsize=7.2cm
\centerline{
\epsfbox{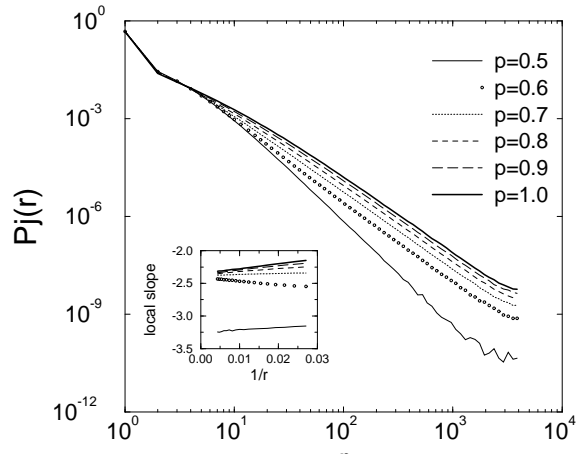}}
\caption{Stationary probability $P_J(r)$ for the peripheral square model with different
values of $p$. Inset: Local slope {\sl vs.} $1/r$. (Note that the slope - exponent $\pi$ -
approaches $-\pi_A=-2.401$ in the limit $r\to \infty$ in all cases with $p\neq 0.5$.)}
\label{fig8}
\end{figure}

\end{document}